\documentclass{article}
\usepackage{import}
\usepackage{color,soul}
\usepackage{placeins}
\usepackage{float}

\usepackage{moreverb,url}
\usepackage[colorlinks=true, linkcolor=black, citecolor=black, urlcolor=black]{hyperref}
\usepackage{bigfoot}
\usepackage{rotating}
\usepackage{graphicx}
\usepackage{booktabs}
\usepackage{pdflscape}
\usepackage{adjustbox}
\usepackage{rotating}
\usepackage{amsmath}  
\usepackage{breqn}
\usepackage{multirow}
\usepackage{float}
\usepackage{tabularx}

\usepackage[sort&compress,numbers]{natbib}
\newcommand\BibTeX{{\rmfamily B\kern-.05em \textsc{i\kern-.025em b}\kern-.08em
T\kern-.1667em\lower.7ex\hbox{E}\kern-.125emX}}

\title{Politics and Propaganda on Social Media: \\How Twitter and Meta Moderate State-Linked Information Operations}

\author{
  Nihat Mugurtay\textsuperscript{1}, 
  Umut Duygu\textsuperscript{2}, 
  Onur Varol\textsuperscript{1,3,*}
}

\date{
    \begin{small}
    \textsuperscript{1}Faculty of Engineering and Natural Sciences, Sabanci University \\ 
    \textsuperscript{2}Sabanci University, Faculty of Arts and Social Sciences, Sabanci University \\ 
    \textsuperscript{3}Center of Excellence in Data Analytics, Sabanci University \\
    \textsuperscript{*}Corresponding author: \texttt{onur.varol@sabanciuniv.edu}
    \end{small}
}

\begin{document}

\maketitle

\begin{abstract}

Why do Social Media Corporations (SMCs) engage in state-linked information operations? Social media can significantly influence the global political landscape, allowing governments and other political entities to engage in concerted information operations, shaping or manipulating domestic and foreign political agendas. In response to state-linked political manipulation tactics on social media, Twitter and Meta carried out take-down operations against propaganda networks, accusing them of interfering foreign elections, organizing disinformation campaigns, manipulating political debates and many other issues. This research investigates the two SMCs’ policy orientation to explain which factors can affect these two companies' reaction against state-linked information operations. We find that good governance indicators such as democracy are significant elements of SMCs' country-focus. This article also examines whether Meta and Twitter's attention to political regime characteristics is influenced by international political alignments. This research illuminates recent trends in SMCs' take-down operations and illuminating interplay between geopolitics and domestic regime characteristics.
\end{abstract}

\maketitle

\newpage
\section*{Introduction}
Political actors are using social media platform to campaign for themselves and influence political agendas in their countries or international relations. Online platforms have become tools for social and political contestation, public opinion, electoral manipulation, disinformation, inauthentic behavior, smear campaigns, and various other forms of online political activities \cite{R13,R14,R15,R16,R17,R18,R19,R20,R21}. 
A number of international and domestic measures have been taken in response to these state-linked concerted actions. SMCs are requested to provide information about activities on their platforms and mitigation plans by governments and platform users. Since 2018, Twitter and Meta have initiated numerous take-down operations against state-linked coordinated activities across the globe. SMCs had various reasons for paying attention to specific networks across different countries: interference in foreign elections, suppression of political opposition, government-affiliated fake news and social bot accounts, and propaganda aimed at legitimizing authoritarian regimes. We examine the premises of Twitter's and Meta's response to state-linked information operations. We explore the politics of social media take-downs originating in many countries, by first exploring the proposed logic of social media take-downs and then using a regression analysis to capture a more delicate analysis of multiple variables, we explore the politics of social media take-downs.

Social media provides a useful modality of public diplomacy, and governments prioritize their digital existence as a part of their national priorities \cite{R79}. At this point, there emerge a complex and interesting phenomenon regarding two American SMCs' response to state-linked information operations. The procedure of taking-down users from specific countries does not merely involve a simple content moderation, but also involves a very close attention to coordinated and state-linked propaganda channels that disseminate political disinformation and government propaganda. Scrutinizing multiple users, groups and companies from different countries, the two SMCs detect concerted political actions which are linked to the governments. This action is directly political in the sense that they focus on government-affiliated entities and decide when, where, and what to moderate. These state-linked actors can be governments, government-related social media companies, political parties, military, security branches of government, individual politicians, municipalities, military, and paramilitary groups \cite{R41}. Therefore, both Twitter and Meta use their own assessment mechanism, regardless of countries' or political regions' \textit{de-jure} framework. When it comes to attempting to take down concerted state-affiliated networks, SMCs have a certain amount of discretion in selecting when, where, and who to target. 

In addition, revenue extraction is also vital for investing in other countries. Giving the fact that content moderation of these companies can be affected from their commercial interests \cite{R69, R70}, conducting state-linked disinformation take downs against governmental entities and private enterprises might result in disrupting their relations with these governments. 
This becomes more salient with the existence authoritarian challenge across the world, where SMCs establish local partnerships, helping Meta and Twitter to generate more wealth by locational advantages. 
Social media giants need to establish good relations with large economies and populations (users) to extract more revenues. On the other hand, democratic countries as well are also subject to coordinated inauthentic activities, which is sometimes initiated by pro-US actors \cite{R80}. Therefore, SMCs also respond to state-linked information operations in the Western Hemisphere, where their main headquarters and decision-making mechanism are located. In this sense, two US-based social media giants are taking action on multiple fronts -in a polarizing geopolitical environment- in response to coordinated-political and state-linked operations. This multi-faceted political, economic, and international dynamics create a puzzle to uncover which factors are significant for take-down policies. 
For instance, Twitter's last report on state-linked information operations was announced just prior to Elon Musk's acquisition of the company turning it into a private entity. 
During the acquisition topics relevant to platform safety such as social bots and misinformation were raised for discussion \cite{varol2023should,rohlinger2023does}.
Couple months later, Musk laid off the team responsible for investigating misinformation campaigns on the platform. Consequently, our research covers the period between 2018 and 2022, beginning with the SMCs' reaction against state-linked information operations and Musk's acquisition of Twitter.

We examine the factors that can be associated with the take down of Meta and Twitter by scrutinizing the states associated with these networks. Domestic political configuration, international politics, and SMCs' focus on revenue extraction from these countries can be significant factors of their reaction against state-linked operations. Therefore, we focus on regime types, countries' international political positioning, and their potential for revenue extraction. 
At this point, we underline that the two American SMCs' reaction against state-linked information operations exist in a complex international environment. Cyberspace and information operations exist with the increasing salience geopolitical setting \cite{R78}. 
We uncover which factors can explain take-down policies of these corporations. 

Building on this puzzle, the challenges of SMCs' political action become more diverse and contentious because these companies deal with political spectrum in multiple countries. This point is significant to show similarities and differences between the two social media giants. In our initial analysis, we find significant parallel and diverging patterns in terms of taking down accounts from specific countries. In this sense, our research scrutinizes Twitter and Meta's take downs in two ways: First, we analyze Twitter and Meta's take-down operations by utilizing an exploratory analysis on countries involved in these operations. We provide a detailed information on why they select specific countries also by analyzing their general discourse and announcements. As a second step, we offer different regression models to determine which variables explains the take downs. We test multiple hypotheses regarding democracy, political stability and violence, and economic indicators. 

\section*{Background on Information Operations}

The actions taken by private SMCs against perilous agents are not a novel phenomenon, yet their policies towards state-linked information operations are relatively new. The literature on SMCs' reaction against inauthentic behavior and online manipulation are mostly associated with content moderation. This is primarily related to insulting, harassment, hate speech, harmful or illegal contents which are also mostly within the scope of countries' domestic legal framework \cite{R67}. Debates on freedom expression \cite{R68}, SMCs' attention for commercial interests \cite{R69,R70}, or regulating content moderation with constitutional and democratic values \cite{R71,varol2016spatiotemporal} have been significant topics for SMCs. Indeed, SMCs' action against governments' information operations go beyond the issues of content moderation.

A state-linked propaganda campaign involves strategic efforts to sway the opinions of targeted groups through a variety of communications tactics with the purpose of achieving a specific outcome \cite{R1,R2,R24}. Coordinated inauthentic behavior on social media encompasses political, social, and psychological dimensions \cite{facebook2018removing}. Individuals often persistently adhere to false belief \cite{R63,R64} and political actors tend to take the advantage of this misinformation environment. Disinformation proliferation has manifested a significant increase, transcending national boundaries and becoming a more globalized concern \cite{R3,R66}. Within this realm, state-linked propaganda as a political communication tool are getting more attention \cite{R32}. Through the involvement of organized groups, online platforms have become an increasingly important component of public diplomacy. 
Disinformation comes into play when misinformation, polarization, fake news or conspiracy theories are deliberately transformed through intentional interventions by governments, political parties, automated bots, coordinated actions by citizens, and internet trolls\cite{R4,R22,R23,R24,R25,R27,R30,R33,R36,R38,R52,R59}. Indeed, particularly bots or concerted disinformation operations are very useful to disseminate low-credible inauthentic content \cite{varol2018deception,R37,ferrara2016rise,varol2017online,varol2020journalists}. Social and political turmoils \cite{R34} \textendash including electoral periods\textendash  are permissive processes for shaping public discourse via disinformation \cite{R39,R47,R55}. In such a disinformation environment, government-run inauthentic social media campaigns grown in response to the mass political mobilizations, and often intend to manipulate elections \cite{R5,R6,R9,R20,R21}.

Twitter and Meta can take-down users from different countries based on inter-state or intra-state factors \cite{R8}. These two modes of concerted action by perilous networks has multiple modalities.
Within the context of intra-state objectives, state-affiliated domestic cyber and information maneuvers are frequently tied to governmental strategies aimed at manipulating narratives to gain support from domestic audience. However, in the international arena, foreign governments use disinformation strategies and many online manipulation tactics to influence geopolitical or foreign domestic processes. Digital sphere can also be an arena of contestation between countries such as China and the United States \cite{R65}, and government-linked inauthentic behavior might target other nations and foreign elections \cite{R31, R66}. In both domestic and international realm, fake news, trolls, disinformation campaigns, conspiracy theories and coordinated authentic or inauthentic actions are the subject matters of these state-linked information operations \cite{R7,R10,R11}. With existence of strong authoritarian countries, the issue of disinformation becomes more interesting to investigate \cite{R42}. In addition, democratic countries also have become an arena of disinformation contest \cite{R44,R61,R80}. This makes state-linked information operations more complex because having alternative source of information \cite{R51} (i.e, social media) is crucial point for democratic regimes. SMCs react against these political manipulation tactics. Focusing on which countries more likely to get attention from SMCs is worthwhile endeavor to illuminate similar or varying patterns. Deciding on which country to focus on against state-linked operations entails a in-house consistent policy-harmonization. Therefore, we also focus on  countries' political orientation towards the United States and other great powers. In this sense, we also search for the factors that might dilute regimes characteristics with geopolitical orientations. At least, we can suggest that the politics of multinationals' matter with the existing polarizing international environment \cite{R77,R78,R79,R80,R81,R82}.

Our research is the first in terms of conducting multi-platform analysis on \textit{state-linked information operations}. We inquiry the political moment transcending a comparison of content moderation policies. By this way, we manifest the trends in take-down strategies. Detecting take-downs and selection specific groups in different countries entail a policy-vision that illuminate different trends for SMCs. In this research, our descriptive and inferential research design uncover these trends. Overall, estimating possible premises of SMCs' state-linked information operations is a substantial contribution to the literature on SMCs' policy outlook. 

\section*{The Politics of Twitter and Meta's Information Operations}

\begin{figure*}[t!]
\centering
\includegraphics[width=\linewidth]{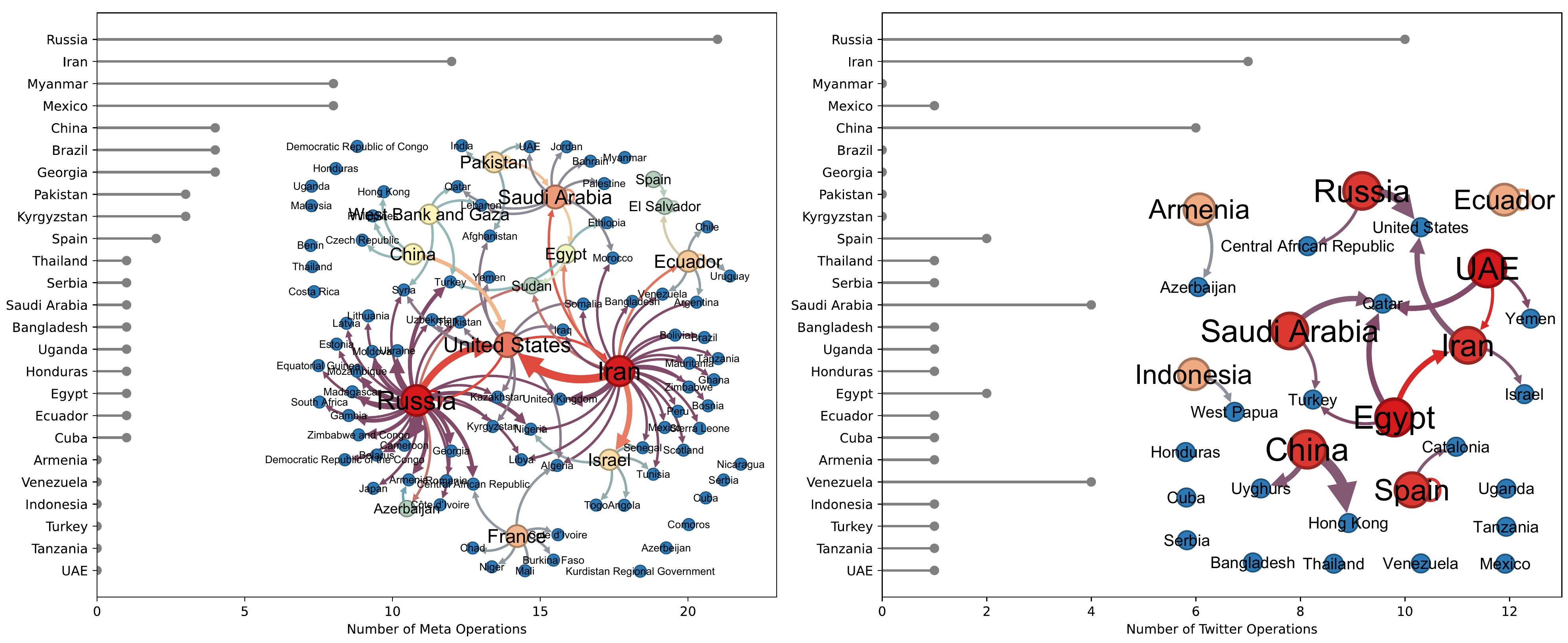}
\caption{\textbf{Take-down operations statistics and country interaction networks for different platforms.} Countries listed by the number of take-down operations held in their country. Discrepancies for Meta (left) and Twitter (right) can be seen by comparing the ranks and the numbers of operations. Networks of interactions are also presented between countries (nodes) and operations (edges from targeting to targeted country). Domestic information operations represented wth self-loops and some of the isolated nodes corresponds to these countries. Node colors represents the out-strengths to measure activities of these countries.}
\label{fig:takedown-stats}
\end{figure*}

Twitter began to disclose its actions against state-linked information operations after 2018. We should also note that SMCs can outsource their counter-action against state-linked information operations. For instance, Stanford Observatory, the institution Twitter cooperates for take-downs, was mentioned in multiple Twitter reports explaining the logic of state-linked information operations. Meta also initially mentioned their partnership with Atlantic Council's Digital Forensic Lab (DFRLab) in taking-down state-linked political networks.
Initially, users from a selected number of countries were taken-down for their government-related coordinated behavior. It is crucial to understand that the motivations behind these take downs vary across countries, as each government engages in specific coordinated activities to shape their social and political agenda. Twitter justifies these take-downs based on both domestic and foreign (inter-state) factors \cite{R8}. Domestically, Twitter has taken action against accounts involved in government propaganda against opposition or ethnic minorities. Internationally, it cites interference in elections on social media as a primary concern \cite{R8}. For instance, Twitter’s action against Iran-linked information operations primarily focused on accounts who allegedly dealt with the US presidential election. In some other cases, Twitter closures are associated with domestic or ethnic issues such as Indonesia's action against West Papua, the 2017 Catalonia's independence referendum. In total Twitter focused on 22 countries around the globe as shown in the supplementary information (SI:Table-1). 
In Figure \ref{fig:takedown-stats}, we mapped out top countries that Twitter and Meta's take action against state-linked information operations. For example, Twitter has taken down accounts from Iran and Russia due to their alleged interference in the international affairs and domestic politics of the United States. Also, it took down accounts originating in China as they involve in Hong Kong and other regional issues. In some other cases, Russia's Internet Research Agency (IRA) conduct operations in Ghana and Nigeria, where suspended accounts dealt with mostly race and civil rights \cite{hern2020russian}. 
These operations are conducted by opening accounts in many African countries. Twitter were able to detect some of these accounts that are working for Russian interests, which are directly affiliated with Russia's IRA. Saudi Arabia, Turkey, and Venezuela are other examples for Twitter platform due to their alleged state-linked information operations. 
For each platform, we built networks mapping of all information operations and their respective target countries through directed edges. In Figure \ref{fig:takedown-stats}(right) illustrates that Twitter has primarily concentrated on perilous networks originating from Saudi Arabia, China, Egypt, Iran, UAE, and Russia. Overall, Twitter has taken measures against perilous networks from 22 different countries, leading to the suspension of several of these networks. However the network constructed from the information operations on Meta suggests Iran and Russia are primary actors and mainly targeting USA.

\begin{figure*}[t!]
\centering
\includegraphics[width=\linewidth]{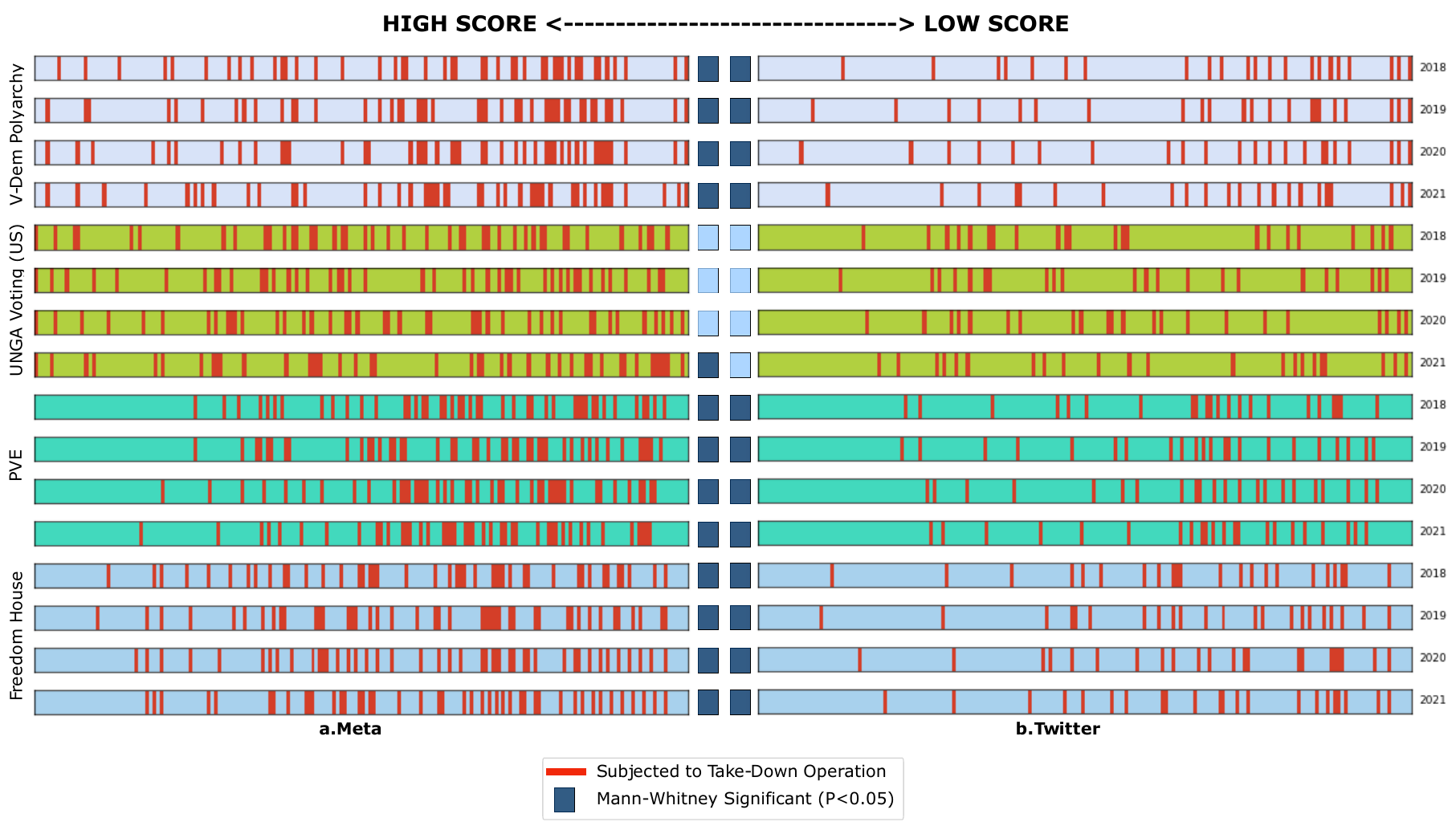}
\caption{\textbf{Positions of countries in different measures.} Ranking of countries taken down by Meta (left) and Twitter (right) presented as red bars. Each row presents different state characteristics scores for years between 2018 and 2021.}
\label{fig:country-ranking}
\end{figure*}

When compared to Twitter's reaction against state-linked information operations, Meta's Coordinated Inauthentic Behavior (CIB) policy focus on various countries in many different regions. In total, we collected information about taken-down perilous networks from 48 countries by Meta, which is significantly more than reported numbers for Twitter. The corporation mainly focuses on concerted users from specific countries claiming they violate the Meta's misrepresentation policy. Different from Twitter, Meta announces both state-linked inauthentic networks and other coordinated political activities under one harmonized framework. Coordinated inauthentic activities driven by commercial interests have also been announced by the Meta. Looking at the content of the Meta suspensions, we also encounter many government-related (executive, military, state-linked companies, municipalities) actions examined by the platform. Meta -for a while- used a specific take-down policy called, \textit{Foreign and Government Inauthentic Behavior (FGI)}, aiming to take-down coordinated activities by such state-linked information operations. However, FGI does not cover all years between 2018 and 2022. Therefore, we manually checked each announcement and document related to CIB and created a full list of Meta's action against \textit{state-linked} information operations. We also discovered that even some accounts publish mainly non-political content about beauty and physical well-being, the Meta detected that these pages are in affiliation with Myanmar military \cite{facebook2018removingMyanmar}. 
The platform suggests that celebrity and beauty pages are useful places to boost state-affiliated networks' impact. This operation against Myanmar-linked accounts has been explained by Meta's emphasis on human rights violations and oppression in the country. In some other instances, Meta revealed a significant portion of inauthentic networks from Russia, and detected that a group of people disguised themselves as journalists and disseminate content about Ukraine and Crimea. The platform claims that this suspended network was in relation with the Russian military. In one example, Meta detected a network affiliated with Islamic Republic of Iran's Press TV, the main broadcasting apparatus of Iran \citep{meta2018coordinatedInauthenticBehavior}.
Meta also pays attention to de-facto or de-jure sub-state entities such as West Bank and Gaza and Kurdistan Regional Government. 

We also have some complex cases, where Meta take downs specific non-state actors such as Muslim Brotherhood , which is an Islamist organization known by their influence on various Islamist movements from charitable organizations to political parties.
Accounts originated in Egypt, Turkey, and Morocco have been taken down by Meta accusing them of committing coordinated information campaigns \citep{meta2020octoberCIBReport}.
In a different particular case, Meta closed accounts from Albania, where a group of Iranians exiled in the country committed coordinated inauthentic behavior \cite{meta2021marchCIBReport}. 
In such complex circumstances, we omitted such governments as targeted by the two SMCs. It is because Turkey, Morocco, and Albania governments were not designated as a part of state-linked information operations by the two SMCs directly. In other words, we coded these countries as being targeted by Twitter and META, if these two SMCs find out a state-linked information networks. We curated a complete picture of Meta's take-down practice by detecting actual state-linked information operations. This entailed an extra layer of hardness to compile Meta's data in detail. In the Western hemisphere, the Meta also carried out its operations against accounts originated countries like Canada and France. In another example, Meta suggested that their teams find a coordinated inauthentic behavior in relation with the French military. At the first sight, as shown in Table 1 in the SI, Meta's operations seem geographically diverse and uncover coordinated and state-linked accounts based on multiple state entity including military, political party, municipality, and any other governmental entity. 

Due to the fact that oscillating regimes towards authoritarianism can oppress their domestic media and find alternative ways of manipulating audience \cite{R12}, both Meta and Twitter appeals to \textit{human rights} and some other governance-related humanitarian emphasis when they announce their suspensions. Second, Twitter and Meta suspend accounts where they see a foreign intervention in other countries' domestic or regional politics. This creates a puzzle regarding their stance based on these premises because there is an important variation in their country-focus list as seen in Figure \ref{fig:takedown-stats}. This also puts significant questions about their state-linked operations since we expect a more harmonized approach due to the global nature of such coordinated misinformation campaigns. 

To describe Meta and Twitter's practice against state-linked information operations, we ranked countries based on different political variables. In Figure \ref{fig:country-ranking}, we present different metrics to provide a basis for comparison, with red bars signifying ranks of these countries that manifest state-linked information operations on that particular scale. 
Each row indicates measures used for comparison such as V-Dem Polyarchy, political similarity with the United States, Political Stability and Absence of Violence (PVE) and Freedom House Index \cite{KaufmannKraay2023,FreedomHouseReport}. 
V-Dem Polyarchy score relies on a robust understanding of democracy, which inspired from Robert Dahl's democratic theory. The measure includes electoral institutions, free press and media and many other civil liberties such as right to protest \citep{CoppedgeEtAl2023VDemOrg}. 
To quantify statistical significance of separation between the countries taken down by SMCs and other by using Mann-Whitney test. This test estimates higher scores if two groups of countries differ in terms of their ranking of the particular metric.
For instance, it is visible that  countries (red bars) clustered around the low score V-Dem's Polyarchy Index, PVE, and Freedom House meaning that both Meta and Twitter more likely to focus on countries with lower democracy and freedom scores. Dark blue rectangles next to ranking plots indicates significant (P-values lower than 0.01) ranking differences by 95 percent confidence level, while light blue denotes non-significance. These scores further reinforce the notion that a country's regime type, freedom scores and political stability are more significant differentiation than geopolitical alignments.  As Figure \ref{fig:country-ranking} illustrates, generally good governance indicators such as democracy and political stability emerge as the most distinguishing factors among when compared to geopolitical stances. However, this initial description needs a more detailed investigation since we still need to detect which premises are the most relevant when compared to others. This is why we employed a regression analysis to capture the politics of Meta and Twitter's information operations. 

Indeed, such an initial observation lead us to key questions: Why does Twitter and Meta specifically focus on certain countries for account take downs, alleging disinformation and manipulation? Are these actions due to these nations engaging in more disinformation compared to others, or are they driven by considerations of human rights, democracy, or geopolitical concerns? Twitter and Meta's justifications encompass political disinformation elements, making it vital to discern which state characteristics factor into these take downs.

\section*{Hypotheses and Data}

Meta and Twitter -as being American multinational SMCs- are giant commercial enterprises that often have complicated relationship with the authoritarian governments and different regime types \cite{R75,R76}. 
In different circumstances, authoritarian countries offer permissive or difficult conditions for multinational international investments. However, this relationship become more salient due to increasing importance of social media as an alternative source of information. 
Fake news and online manipulation are shown to be effective weapons to undermine democratic ideals \cite{R26}. Both Twitter and Meta mention governance-related issues such as democracy, elections, justice, and human rights in their reports, and we expect from both that authoritarian regimes are on the top of their priorities in global information landscape \cite{R44,R61}. 
State-linked information operations are also manipulation tools for diverting both domestic and foreign attention. China, Iran, and Russia are notable instances of information operations. Particularly, their online intervention in regional and electoral agendas are well pronounced by the SMCs. More authoritarian governments can use social media as a public diplomacy tool to divert the public's attention from their oppression. However, we also have other non-authoritarian countries that were scrutinized by SMCs. It shows that information operations are not unique to only autocratic governments. However, we need to have a clear measure if SMCs focus on authoritarian countries more than democratic governments. In order to identify possible differences between Meta and Twitter's emphasis on authoritarian inclinations, we operationalize different political and governance-related variables. \\

\textbf{H1.} Social media platforms are more likely to take actions against state-linked information operations if they originate in more authoritarian countries.\\

The geographical variation of state-linked information operations may not be explained solely by regime oscillations and companies' reactions against more authoritarian regimes. However, not all authoritarian countries have the same level of repressiveness or political instability. From another perspective, there can be more democratic countries with higher levels of political instability and violence. In democratic countries with higher political instability, political parties and different segments of state bureaucracy can use disinformation and other online manipulation tactics to convey their propaganda. These less authoritarian countries can be more open to state-linked information operations including countries such as Mexico, Nigeria, and Ukraine. Alternatively, we can expect that non-democratic countries with more stability focus on their people's well-being, while some authoritarian countries use their force to oppress domestic dissent. The Meta documents clearly the emphasize on human-rights violations committed by oppressive regimes. These regimes are primarily target the physical integrity of their own population. Large SMCs may take down accounts originating in countries with less political stability, reflecting concerns about domestic violence and other relevant issues. To examine this hypothesis, we used the ``absence of violence'' variable from the World Bank's Good Governance Indicators.\\  

\textbf{H2.} Social media platforms carry out take-down actions against users where political violence and political instability more likely to take place.\\

\begin{figure*}
\centering
\includegraphics[width=\linewidth]{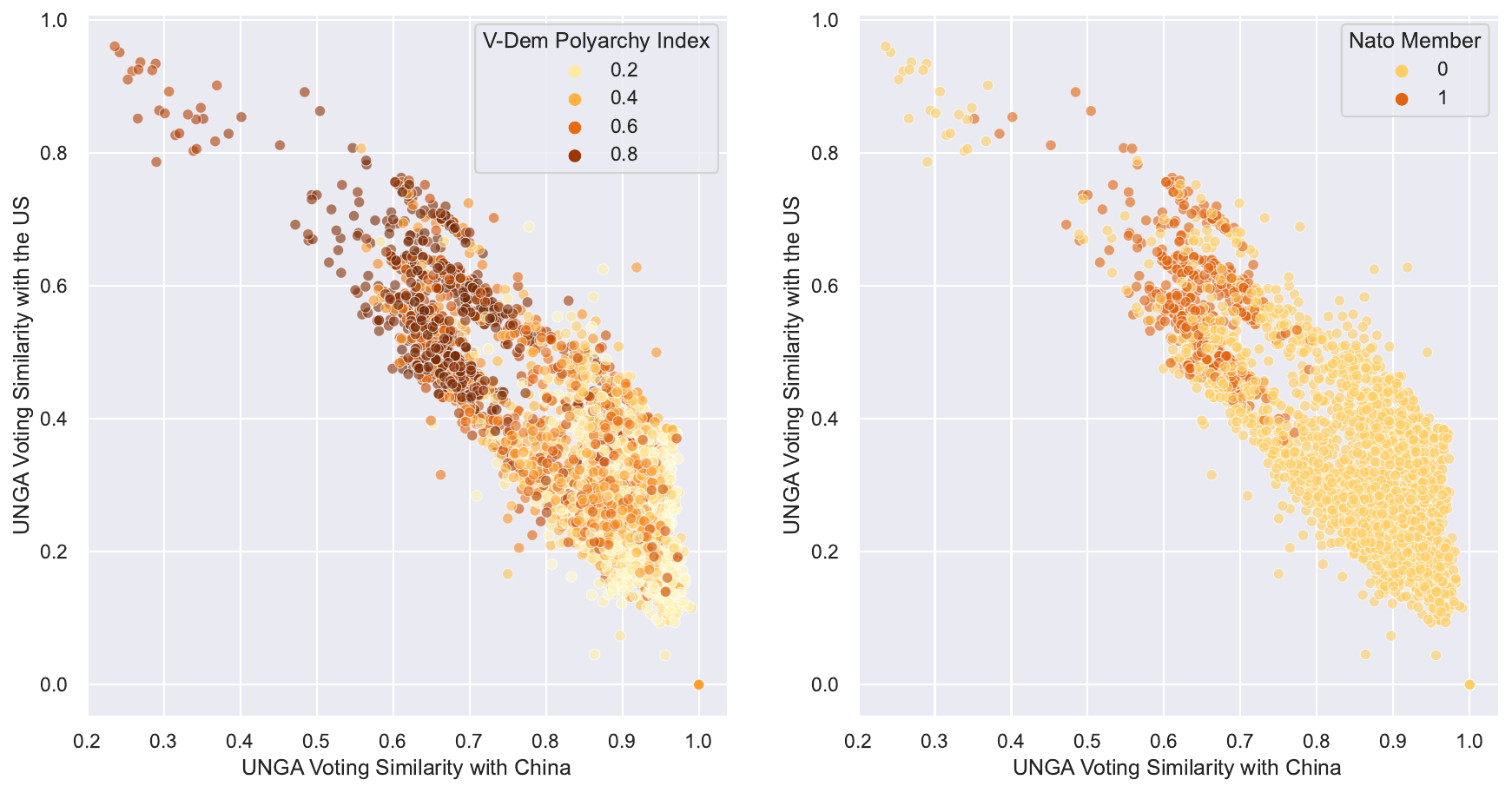}
\caption{\textbf{UNGA Voting similarities of countries to US and China.} UNGA voting similarity shows negative association between United States and China. More democratic countries in V-Dem Polyarchy index have similar voting pattern to US (left) and Nato members also more likely to align with preferences of US (right).}
\label{fig:unga-sim}
\end{figure*}

In addition to governance indicators, Cyberspace has been an arena of international relations \cite{R77}. This space creates a policy pressure on the United States to be more active against state-linked information operations \cite{R81}. In such a geopolitical environment, some countries are following the footsteps of great powers and express changing levels of similarity in political outcomes \cite{R29}. We note that both Meta and Twitter are multinational corporations. In other words, they can establish local partnerships with other parties, outsource some of their operations to third parties and open branches in multiple geographies. They can also be affected from host countries' political conditions \cite{R73}. Although they tend to globalize their commercial activities, they are American multinational SMCs, meaning that their headquarters and decision-making process take place in the United States. Multinational SMCs are not exempt from their country-of-origin characteristics and are affected by the international geopolitical agenda of their home countries \cite{R72, R74}. Therefore, international alignments might be an influential factor for SMCs. United Nations General Assembly (UNGA) voting patterns is a useful tool to test if countries have a similar mindset or policy-orientation in international affairs. It should be noted that UNGA Similarity does not directly manifest the strong de-facto or de-jure political alliances, but it shows how countries are similar to each other in international politics. It gives a picture of how countries cast their votes similarly at the international level. We investigate if multinationals carry out their take-down operations focusing on the countries with less UNGA voting similarity or not. With this regard, we expect that international political similarity is a significant factor of take-down policies of MNCs.

\textbf{H3.}  Users originating from countries with higher political similarity to the United States are less likely to have their accounts taken down by Meta and Twitter.

Foreign policy and geopolitical concerns are becoming more substantial for multinational companies \cite{R82}. Concerns for human rights and democracy exist in such a geopolitical setting \cite{R28}. SMCs operate under the pressure of a geopolitical and polarized world where authoritarian countries emerge as contending actors, and social media operations are part of great power politics \cite{R78}. In this sense, we investigate if Twitter and Meta's reaction against more authoritarian counties are moderated by these countries' political similarity with the United States or China. To address this, we concentrate on a possible conditional effects illuminating if good governance indicators (democracy and absence of violence) interact with great powers' positioning in international issues. We take this similarity as a reflection of general political like-mindedness in the international system (not as a reflection of de-jure or de-facto political alliances). Indeed, these similarity matters because great powers usually manifest a clear divergence from each other. To illustrate, as seen in the Figure \ref{fig:unga-sim}, if a country has a high voting similarity with the United States, then this country more likely to have a low voting similarity with China between 2000 and 2021. Because UNGA voting patterns show opposite trends for great powers, we are confident to use UNGA Voting similarity to measure countries like-mindedness in the political realm. In Biden-Harris National Security Strategy, China is also defined as ``the only competitor'' that has a potential to transform the international political and economic landscape \cite{NationalSecurityStrategy2022}. We also suggest that China is labelled as the geostrategic rival by the US authorities. It is also visible that NATO member countries have relatively higher UNGA voting similarity with the United States, indicating how UNGA voting similarity reflects a systemic understanding. Therefore, using UNGA voting similarity with China will be appropriate to see if it follows our findings on the US case. 

There is a geopolitical issue here when it comes to democratic backsliding, authoritarian regimes and multinational corporations. Moreover, Figure \ref{fig:unga-sim} (left) shows that higher V-Dem Polyarchy scores are associated with more political similarity with the United States. Therefore, UNGA agreement scores can not only reflect geopolitical orientations but also a partnership of democratic countries at the international level.
Multinational SMCs are not immune to these global trends since economic blocks, regionalization, and degrading globalization are substantial challenges. For instance, home countries call some of their high-tech companies to return back by channeling significant amount of incentives. 
In some other examples, geopolitical changes divert capital from rouge states to more politically like-minded countries. Therefore, as American multinationals, Meta and Twitter might more openly examine state-linked information operations in countries that have a low level of political similarity with the United States. Here, we do not offer a causal link but we expect a negative association between SMCs' take-down operations and UNGA voting similarity with the United States.\\

\textbf{H4.} More authoritarian or oppressive countries with higher political similarity to the United States are less likely to be scrutinized by Meta and Twitter.\\

An important indicator of SMC investments is the number of users and revenue extracted from a country. SMCs' global outreach is a useful tool to establish local partnerships and extract more revenue. 
In simpler terms, large populations are essential for growing markets and they can be the focus of big SMCs. In similar fashion, we observe a strong correlation between the population and the number of social media users (SI. Fig-2). Therefore, population can be used to assess the global outreach and power of SMCs. 

We also suggest that the economic power of countries may act as a deterrent for SMCs not to interfere with state-linked operations. A social media company makes its profit globally as a multinational actor investing in variety of countries across the globe. They form alliances with local partners for a variety of profit-making purposes. Consequently, wealthier countries become attractive economic markets due to increasing business, advertising, and their high demand for technology. For example, focusing on India and Brazil, and other countries with emerging markets, SMCs specifically tailor their services to the domestic audience in these regions. 

However, users' marginal economic contribution also appears to be more important than the revenue opportunities provided by wealthier countries based on advertisements, marketing, and other commercial activities. In particular, average revenue per user (ARPU), which is a key indicator of the profitability, indicates whether or not SMCs will generate a greater amount of revenue from one additional user. 
Generally, revenue generation is centered in regions with high ARPUs. Currently, we are using regional estimates provided by Meta and Twitter \cite{FacebookStatsDataTrends2023}.
Taking down state-linked accounts from countries with higher ARPUs could be avoided by Meta and Twitter. In contrast, we do not have ARPU for all countries. Thus, since ARPU is directly related to GDP per capita, we use that as proxy for ARPU and assumes adding one more user from countries with a higher GDP per capita will result in greater revenue for SMCs. Taking down users from such countries can be costly for SMCs.\\

\textbf{H5.} Twitter and Meta are less likely to focus on countries with larger populations and higher GDP per capita.

\section*{Research Design}

To address the hypothesis introduced in the previous session, we use a binary dependent variable model, and operationalize different logistic regression models.
These models capture factors that are associated with Twitter and Meta's take-downs in particular countries. 
We first used pooled logistic models to observe between-unit trends in time with clustered standard errors (CSE). Using CSE at the country level in pooled models overcome the problems associated with serial correlation and heteroskedasticity. The first equation shows our first estimation for non-interaction models. The second equation shows an example of interaction models of good governance indicators and political similarity with the United States. For more robustness, we also used linear probability models to demonstrate if our models follow a similar output pattern as seen in tables 2 and 3 in the supplementary material (SI:Table 2 and 3).

Our regression model in Table \ref{table:regressions-pooled} covers the years between 2018 and 2022, since Twitter reported state-linked operations until 2022. To analyze and illustrate the differences between countries, we marked countries as $1$ for all years between 2018 and 2022 if they experienced a take-down operation by Twitter or Meta during these years. 
This also touches the point that Meta and Twitter handled their reaction to state-linked information operations. 
They follow these coordinated networks for a long time to see how they operate \cite{meta2018coordinatedInauthenticBehavior}. Therefore, scrutinizing a perilous network can take even two or more years. Even this might be the case, we used also cross-sectional logistic models with robust standard errors to capture between-unit variance more clearly. In these models, our continuous predictors are four-year averages between 2018 and 2022. 

To test our hypotheses, we curated a dataset from Twitter and Meta's official web pages. We find out that not all Meta's politics-related focus is addressing state-linked information operations. We checked each announcement to see if Meta builds a linkage between inauthentic networks and governments. If they underline these linkages, then we coded this operation as state-linked. 

\begin{dmath}
\log \left( \frac{\text{P(T=1)}}{1 - \text{P(T=1)}} \right) = \beta_0 + \beta_1 \text{V-Dem Polyarchy} + \beta_2 \text{UNGA(w/USA)} + \beta_3 \text{Political Stability} + \beta_4 \text{GDP per capita(log)} + \beta_5 \text{Population(log)} + \epsilon
\end{dmath}

\begin{dmath}
\log \left( \frac{\text{P(T=1)}}{1 - \text{P(T=1)}} \right) = \beta_0 + \beta_1 \text{V-Dem Polyarchy} + \beta_2 \text{UNGA(w/USA)} + \beta_3\text{V-Dem Polyarchy}\times\text{UNGA(w/USA)}+ \beta_4 \text{Political Stability} + \beta_5 \text{GDP Per Capita(log)} + \beta_6 \text{Population(log)} + \epsilon
\end{dmath}

To explore state-characteristics, we provide general datasets and their scores encompassing regime type, freedoms, corruption, and UN-Voting similarity with the United States. For each model, we used V-Dem Polyarchy Score \cite{CoppedgeEtAl2023VDemOrg}, Political Stability and Absence of Violence \cite{KaufmannKraay2023},  UN-Voting Similarity estimates \cite{DVN/LEJUQZ_2009}, Gross Domestic Product (GDP), and Population from World Bank. We transformed GDP Per Capita and population with their logarithms in base 10. 
According to the variance inflation factors (VIF) provided in the supplementary material (SI:Table-7), we did not find any multicolliniarity between the V-Dem Polyarchy Index and Political Stability score.

Since, user statistics is another important indicator of SMCs' market potential in a country, we have substituted population data with user statistics and incorporated our regression models, as detailed in supplementary information (SI:Table-5). Although this approach provide more robustness to our findings, we want to point that China, Iran, and Russia (the most scrutinized countries) lacks official user statistics for these platforms. This explains the reasons we have to rely on population data instead of user statistics. 

Regarding the political economy of multinational SMCs, we also add the UNGA Voting Patterns for China in the supplementary material (SI:Table-4). If there is a moderating effect of UNGA Similarity with the United States, then we should at least observe a moderating effect of political similarity with China and Russia. This will provide more robustness if we find such a tendency, reflecting the existing environment for take-down operations. Indeed, our main interest here is to show how UNGA Voting Patterns with the United States change Meta and Twitter's attention for more authoritarian countries. Our exploratory analysis and regression output show significant parallel trends between Twitter and Meta. 
\\

\begin{table*}[!]
\centering
\caption{Regression Results (Logit Estimates, Pooled Time-Series)}
\label{table:regressions-pooled}
\begin{adjustbox}{width=1.0\textwidth}
\begin{tabular}{l*{10}{c}}
\hline\hline
                    &\multicolumn{1}{c}{(1)}&\multicolumn{1}{c}{(2)}&\multicolumn{1}{c}{(3)}&\multicolumn{1}{c}{(4)}&\multicolumn{1}{c}{(5)}&\multicolumn{1}{c}{(6)}&\multicolumn{1}{c}{(7)}&\multicolumn{1}{c}{(8)}&\multicolumn{1}{c}{(9)}&\multicolumn{1}{c}{(10)}\\
                    &\multicolumn{1}{c}{Meta}&\multicolumn{1}{c}{Twitter}&\multicolumn{1}{c}{Meta}&\multicolumn{1}{c}{Twitter}&\multicolumn{1}{c}{Meta}&\multicolumn{1}{c}{Twitter}&\multicolumn{1}{c}{Meta}&\multicolumn{1}{c}{Twitter}&\multicolumn{1}{c}{Meta}&\multicolumn{1}{c}{Twitter}\\
                   
\hline
               &               &               &               &               &               &               &               &               &               &               \\
V-Dem Polyarchy     &      -2.146***&      -3.540***&      -1.344   &      -3.631***&      -2.316** &      -3.731** &      -0.803   &       1.641   &      -1.573   &      -3.108** \\
                    &      (0.72)   &      (1.19)   &      (0.90)   &      (1.33)   &      (1.08)   &      (1.59)   &      (2.33)   &      (3.38)   &      (1.07)   &      (1.57)   \\
PVE                 &               &               &      -0.409*  &       0.059   &      -0.183   &       0.198   &      -0.192   &       0.244   &       1.637** &       1.332** \\
                    &               &               &      (0.21)   &      (0.22)   &      (0.28)   &      (0.39)   &      (0.29)   &      (0.42)   &      (0.64)   &      (0.64)   \\
UNGA Voting (w/USA) &               &               &               &               &       0.806   &      -2.606   &       4.292   &       9.322   &      -0.022   &      -3.430   \\
                    &               &               &               &               &      (1.88)   &      (2.31)   &      (5.19)   &      (6.90)   &      (1.96)   &      (2.27)   \\
GDP Per Capita (log)&               &               &               &               &       0.216   &       0.577** &       0.257   &       0.655** &       0.335*  &       0.626** \\
                    &               &               &               &               &      (0.18)   &      (0.29)   &      (0.19)   &      (0.29)   &      (0.20)   &      (0.30)   \\
Population (log)    &               &               &               &               &       0.640***&       0.841***&       0.652***&       0.862***&       0.690***&       0.848***\\
                    &               &               &               &               &      (0.16)   &      (0.27)   &      (0.16)   &      (0.27)   &      (0.16)   &      (0.27)   \\
V-Dem Polyarchy $\times$ UNGA Voting (w/USA)&               &               &               &               &               &               &      -5.317   &     -18.908*  &               &               \\
                    &               &               &               &               &               &               &      (7.22)   &     (10.92)   &               &               \\
PVE $\times$ UNGA Voting (w/USA)&               &               &               &               &               &               &               &               &      -6.811***&      -4.262** \\
                    &               &               &               &               &               &               &               &               &      (2.51)   &      (1.80)   \\
Constant            &       0.030   &      -0.482   &      -0.496   &      -0.419   &     -12.649***&     -18.639***&     -14.113***&     -22.723***&     -14.351***&     -19.096***\\
                    &      (0.39)   &      (0.50)   &      (0.49)   &      (0.60)   &      (3.02)   &      (5.09)   &      (3.88)   &      (5.06)   &      (3.30)   &      (5.03)   \\
\hline
Number of clusters  &     168   &     168  &     168   &     168   &     168   &     168   &     168   &     168  &     168   &     168   \\
Number of observations&     672   &     672   &     672   &     672   &     672   &     672   &     672  &     672   &     672   &     672   \\
\hline\hline
\end{tabular}
\end{adjustbox}
\end{table*}

\begin{table*}[!]
\centering
\caption{Regression Results (Logit Estimates, Crosssection)}
\label{table:regressions-crosssection}
\begin{adjustbox}{width=1.0\textwidth}
\begin{tabular}{l*{10}{c}}
\hline\hline
                    &\multicolumn{1}{c}{(1)}&\multicolumn{1}{c}{(2)}&\multicolumn{1}{c}{(3)}&\multicolumn{1}{c}{(4)}&\multicolumn{1}{c}{(5)}&\multicolumn{1}{c}{(6)}&\multicolumn{1}{c}{(7)}&\multicolumn{1}{c}{(8)}&\multicolumn{1}{c}{(9)}&\multicolumn{1}{c}{(10)}\\
                    &\multicolumn{1}{c}{Meta}&\multicolumn{1}{c}{Twitter}&\multicolumn{1}{c}{Meta}&\multicolumn{1}{c}{Twitter}&\multicolumn{1}{c}{Meta}&\multicolumn{1}{c}{Twitter}&\multicolumn{1}{c}{Meta}&\multicolumn{1}{c}{Twitter}&\multicolumn{1}{c}{Meta}&\multicolumn{1}{c}{Twitter}\\
                      \\
\hline
                &               &               &               &               &               &               &               &               &               &               \\
V-Dem Polyarchy     &      -2.271***&      -3.601***&      -1.368   &      -3.691***&      -2.471** &      -3.758** &      -0.796   &       2.033   &      -1.700   &      -3.105*  \\
                    &      (0.73)   &      (1.21)   &      (0.92)   &      (1.35)   &      (1.16)   &      (1.66)   &      (2.55)   &      (3.58)   &      (1.15)   &      (1.64)   \\
PVE                 &               &               &      -0.460** &       0.059   &      -0.195   &       0.201   &      -0.206   &       0.246   &       1.788** &       1.445** \\
                    &               &               &      (0.22)   &      (0.22)   &      (0.30)   &      (0.40)   &      (0.30)   &      (0.44)   &      (0.79)   &      (0.69)   \\
UNGA Voting (w/USA) &               &               &               &               &       1.102   &      -2.625   &       5.057   &      10.712   &       0.321   &      -3.470   \\
                    &               &               &               &               &      (2.07)   &      (2.53)   &      (6.02)   &      (7.65)   &      (2.20)   &      (2.40)   \\
GDP Per Capita (log)&               &               &               &               &       0.191   &       0.588*  &       0.235   &       0.664** &       0.319   &       0.637** \\
                    &               &               &               &               &      (0.18)   &      (0.30)   &      (0.20)   &      (0.30)   &      (0.20)   &      (0.31)   \\
Population (log)    &               &               &               &               &       0.665***&       0.839***&       0.680***&       0.861***&       0.719***&       0.851***\\
                    &               &               &               &               &      (0.16)   &      (0.28)   &      (0.16)   &      (0.28)   &      (0.17)   &      (0.28)   \\
V-Dem Polyarchy $\times$ UNGA Voting (w/USA)&               &               &               &               &               &               &      -5.960   &     -20.669*  &               &               \\
                    &               &               &               &               &               &               &      (8.21)   &     (11.82)   &               &               \\
PVE $\times$ UNGA Voting (w/USA)&               &               &               &               &               &               &               &               &      -7.547** &      -4.658** \\
                    &               &               &               &               &               &               &               &               &      (3.26)   &      (1.88)   \\
Constant            &       0.122   &      -0.455   &      -0.473   &      -0.392   &     -12.867***&     -18.696***&     -14.506***&     -23.134***&     -14.707***&     -19.234***\\
                    &      (0.39)   &      (0.51)   &      (0.50)   &      (0.61)   &      (3.10)   &      (5.16)   &      (4.14)   &      (5.13)   &      (3.45)   &      (5.11)   \\
\hline

Number of observations&     168   &     168   &     168   &     168   &     168  &     168   &     168   &     168   &     168   &     168   \\
\hline\hline
\end{tabular}
\end{adjustbox}
\end{table*}

\begin{figure*}
    
\centering
\includegraphics[width=\linewidth]{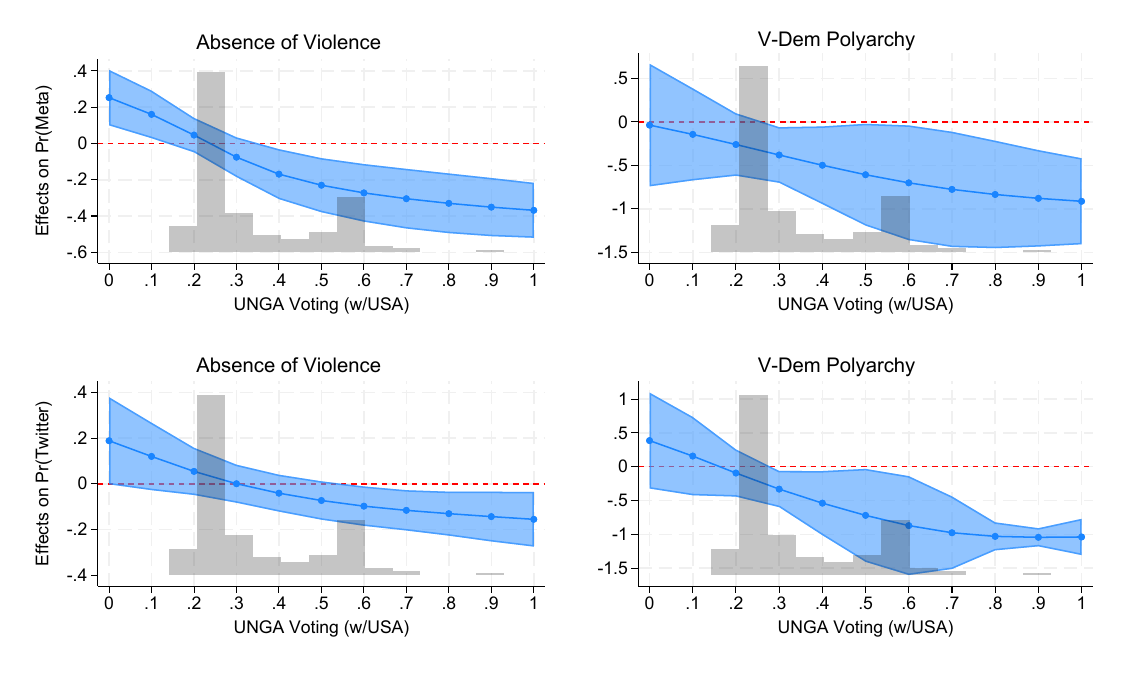}
\caption{Conditional Effects of Polyarchy and Political Stability on Meta and Twitter's Country Focus}
\label{fig:conditional-effects}
\end{figure*}

\section*{Results and Discussion}

Our pooled and cross-sectional logit estimates show significant implications to test hypotheses presented in this work. 
Table \ref{table:regressions-pooled} shows the time-series pooled logit estimates for 2018-2022, while Table \ref{table:regressions-crosssection} is the cross-sectional version of same models. Since H1 and H2 are related to governance-related issues, models 1-4 in both regression tables shows the association between governance indicators and SMCs take-downs. 
We initially assessed whether Twitter and Meta's take-down operations focus more on authoritarian countries. The bi-variate regression model in the first two columns indicates that both are concentrating on less democratic countries. In simpler terms, the likelihood of SMCs directing their actions towards countries rises as V-Dem scores decrease. Although Meta and Twitter may focus on varying numbers of countries, they exhibit a similarity in addressing accounts originating from less democratic nations. This effect is more strong and significant for Twitter. Our analysis in Figure \ref{fig:country-ranking} also reflects this finding. 

Authoritarian countries vary in repressiveness and political instability. Some democracies face higher political instability and violence. Political parties and state bureaucracies in these democracies may use disinformation and online manipulation for propaganda, making them sensitive to state-linked information operations. Models 3 and 4 in both tables reveal that Meta primarily focuses on accounts originating from countries with higher levels of political instability. Conversely, Twitter continues to direct its attention towards countries with more authoritarian tendencies, even after including the political stability variable, as seen in Model 4. After adding other independent variables, we observe that good-governance indicators protect their effect on SMCs' take-downs for columns 5 and 6. On Column 5 and 6 of \ref{table:regressions-crosssection} and \ref{table:regressions-pooled}, we see a significant negative association between Twitter and META's take-downs and countries' democracy score at 95 percent confidence level. 

In summary, the initial six models without interaction terms indicate that governance indicators, particularly play a significant role in the SMCs' country focus vis a vis other variables. Based on this, H1 is corroborated for the first six regression column of Table \ref{table:regressions-crosssection} and \ref{table:regressions-pooled}.

We also put a caution here. Our findings do not suggest that more authoritarian countries commit more state-linked information operations. We estimate our regression models based on what Twitter and META reported. Such operations are not unique to authoritarian countries. For instance, we see META also focuses on United States and France if they detect a state-linked network, meaning that Western democracies also can be focused on by SMCs. If such take-downs happen, it means that democratic countries are also subjects of state-linked information campaigns. However, Meta and Twitter's take-down policies reflect divergences in investigating Western democracies. For instance, Twitter never carried-out its take-down operations in relation with any Western Democracies. This puts a question mark regarding the politics of take-down operations. This brings us the idea behind H3 and H4. We also investigate if there is an underlying geopolitical logic in these two SMCs' information operations. Domestic state regime characteristics may be influenced by the presence of realpolitik in international affairs.

Multinational companies are powerful actors regarding market information and their competitiveness in domestic markets, which complicates their relationship with the sovereign states \cite{nye1974multinational,gu2023data,Lowe2021}. 
Since these companies might be affected by international politics, we add UNGA Voting similarity with the United States. 
Model 5 and 6 in both Table \ref{table:regressions-crosssection} and \ref{table:regressions-pooled} manifest non-interaction models with other independent variables including UNGA Voting (w/USA), Population (log) and GDP Per Capita (log). These analysis yield no significant relationship between SMCs' take-downs and countries' political similarity with the United States. 
In other words, Twitter and Meta take actions for countries regardless of their political like-mindedness with the US at the international level. This finding is valid for our all models, which falsifies H3. These finding is important because these take-down operations also carried out for many reasons including international reasons such as election intervention, anti-American propaganda. Indeed, we do not find a significant effect of political like-mindedness with the United States on the two SMCs' take-down operations. 

Since disinformation landscapes are a complex amalgam of domestic and international politics, we used an interaction variable based on good governance indicators and UNGA Voting similarity with the United States. Models 7-10 reflects this idea focusing on if international political similarity moderates the effect of good-governance indicators. 
In these models, we observe that political similarity with the United States has a moderating effect on good governance indicators, particularly in columns 8, 9 and 10 in both tables. In other words, the effect of democracy and political stability changes with different levels of UNGA Voting (w/USA). Since regression coefficients in logistic models are odd-plotting marginal and conditional effects are useful tool to interpret interaction coefficients. We plotted conditional effect of good governance indicators'  as seen in the Figure \ref{fig:conditional-effects}. All four figures show how the effect of good governance indicators vary for different levels of UNGA Voting Similarity with the US. Subplots on the top row show that the conditional effect of absence of violence and V-Dem democracy scores on Meta's take-downs changes for different levels of UNGA Voting Similarity (w/USA). For Meta, we find that the effect of good governance indicators are more visible for higher values of political similarity with the United States. This is valid for both political stability and democracy scores. For Twitter, the interaction effect is more visible for V-Dem Democracy score (lower right in Fig.\ref{fig:conditional-effects}). But, the same is not valid for absence of violence. This can also be seen in the different models of regression tables, and this conditional effect is more pronounced for higher levels of UNGA Voting (w/USA).  

It is interesting that although political similarity with the United States is not significant based on main effects as seen in models 5-6, it has a interaction effect on good-governance indicators. At this point, we suggest that our models 7-10 corroborates H4. However, we approach this finding with a caution. As seen in Figure \ref{fig:conditional-effects}, we have also included a histogram of UNGA Voting (w/USA) to determine if its distribution is skewed. Its distribution, evident in the four figures, demonstrates a skew towards lower values. Given this, we need to pay attention in interpreting its effects beyond the 0.6 mark, where the data points are notably sparse. While corroborating our H4, we do not offer any causality regarding this moderating effect.  

In addition to good governance indicators and international alignments, we tested whether countries' economic power and population manifest an association between Twitter and Meta's reaction to information operations. As seen in H5, we expected that there can be an economic logic for SMCs, preventing them carrying out their take-down operations. However, it is observed that users originating in countries with larger populations are more likely to be taken down by SMCs, a trend consistent across all regression models for both tables. This important because larger populations imply larger user statistics and revenue extraction (SI:Table-6 and Figure-2). Also, we propose that a higher GDP Per Capita represents an additional potential for revenue extraction by users. Contrary to expectations, our analysis did not confirm that SMCs are less inclined to take down accounts originating in countries with higher GDP per capita. As a result, we found no economic rationale underlying the take-down policies of SMCs. Interestingly, we discovered evidence that Meta conducts counter-operations even against the United States, which is the country where it generates the highest revenue \cite{meta2022adversarialThreatReportQ3}. 
This is the opposite of what we expected. It illustrates that while scrutinizing state-linked accounts originating in countries, Meta and Twitter is carrying out their operations against state-linked coordinated activities regardless of countries' economic power.  

These findings can explain how Meta and Twitter are similar or different in terms of their response to state-linked information operations in different countries. While they are more hesitant to different good-governance indicators, political similarity with the US has a moderating effect for both SMCs. We demonstrated that both Twitter and Meta are responsive to countries' good governance indicators such as democracy and political stability, based on pooled logit and cross-sectional estimations. However, we also underline the importance of political similarity with great powers at the international level. Therefore, a demarcation line between authoritarian and democratic countries -regarding SMCs' response to state-linked information operations- can be misleading. This is the point that we observe the possible conditional effect of international alignments on SMCs' action against different states. 

There is a trend towards authoritarianism globally. However, these authoritarian inclinations occur in a multi-polar international landscape. Disinformation and concerted perilous networks can nurture in places when they are protected by a powerful state apparatus such as military or security entities. These networks operate in an international landscape, where geopolitical shifts take place including inter-state and civil-wars. While multinational corporations seek more revenue globally, they also suffer acute or long-term geopolitical crises. Countries such as China, Iran, North Korea and Russia ban access to the two SMCs. The countries are also politically distant from the United States and other western countries. Thus, we can see how autocratic tendencies meddle with international alignments. Our study suggests two possible interpretations regarding the prevalence of state-linked information operations. First, it is conceivable that authoritarian regimes provide a permissive environment for the proliferation of state-linked information operations, which would explain the higher frequency of account take-downs by SMCs. Secondly, it is plausible to consider that SMCs utilize a selective mechanism in their regulatory actions, focusing not only on the oppressive nature of these operations but also on their involvement in international affairs, including interference in foreign elections and domestic politics. Several networks have been targeted for this purpose, including those linked to China, Iran, and Russia. For instance, Meta, through 116 separate announcements on state-linked operations, identified 38 of them as foreign interference. Within this 38 foreign operations, Meta reports that United States was targeted by seven times by other country-linked perilous networks such as Russia, Iran and China. However, Egypt and the UAE-linked networks target multiple countries including Iran(7), Turkey(5) and Qatar(5)\footnote{Number of separate operations in parenthesis}. Based on Twitter's 48 released datasets (separate announcements), United States has been the top country targeted by these state-linked information operations (SI: Table 8). Qatar, Hong Kong and Turkey are other countries as being targeted by state-linked perilous networks. That is, when it comes to target of taken-down networks, we see variations across different SMC platforms. According to Twitter reports and our statistical anaylsis, it pays more attention to domestic-based actions.  

This underscores the complex and dynamic nature of digital geopolitical landscape, where state-linked information operations are not confined to a single strategy or target. It highlights the need for ongoing outlook and adaptive strategies by SMCs to counter these evolving threats and protect the integrity of global information landscapes. It is important to note, however,  we do not provide any anecdotal or qualitative evidence. Indeed, our findings regarding conditional effects show how regime characteristics and international political similarity interact with each other statistically.

All findings above would be challenged or corroborated with with more data coming from SMCs. Each social media or information technology company has different levels of transparency\cite{urman2023transparent}. Twitter offers user-level metadata, which allows researchers to analyze take-down operations in a comprehensive manner. User characteristics and tweets inform us about their possible network, replies, quote tweets and many other user-level traits. However, Meta or Facebook does not provide open access to this type of information. If an in-depth individual-level data were available for both platforms, a deeper understanding of the underlying premises guiding their take-down operations would be possible. In-country and between-country comparisons would be possible and helpful to understand how they detect these networks. Researchers can extend their research on SMCs' take-down policies, which would be an integral part of  transparency in the cyberspace and disinformation landscape.

\section*{Conclusion}

SMCs operate in a challenging global environment. This includes dealing with authoritarian regimes in a world with many powerful nations. The shift away from democracy is and important factor for the spread of misinformation. Social media emerges a key and contested arena for countries to implement their (dis)information strategies. These efforts often aim to bolster their domestic and international agendas. Governments can use this to build support locally or to distort the truth about events to gain international backing, since social media is a common platform for interaction among individuals, political parties, and politicians. Based on Meta and Twitter's reports, this is evident in actions by countries like Russia, Iran, and China and many other countries, where they have a capability to influence foreign elections, create fake news, engage in misleading public diplomacy, and spread misinformation. Therefore, it is crucial for SMCs to develop a coherent approach against these state-linked activities and to fight misinformation through social media. When SMCs address these issues, they act in a political environment because the subjects of their take-down operations are state-linked. This is the main difference between these operations and classical content moderation. 

This study provides significant insights into the behavior of SMCs, specifically for Meta and Twitter. We have found that domestic governance indicators are important for both SMCs. Our findings indicate that SMCs tend to focus on countries with lower democracy scores and higher levels of political instability. This suggests that governance indicators play a crucial role in shaping SMCs' focus, which aligns with the hypothesis that less democratic nations are more likely to experience such operations. 

However, we also questioned if good governance indicators' impact is affected by international political tendencies. In other words, we ask if the impact of good governance indicators is moderated by UNGA Voting Similarity with the U.S. Despite the lack of direct evidence between SMCs' actions and international political like-mindedness, it is interesting to note that we find a modest evidence that the effect of good governance indicators change with different levels of like-mindedness with the United States and China. Considering both internal governance structures and international relations when analyzing SMCs' operational strategies highlights the multifaceted nature of their responses to global information manipulation. The politics of Twitter and Meta's operations matter in such a geopolitical landscape. We can conclude that authoritarian challenges occur in a geopolitical landscape which is a reflection of differentiating geographical orientations in world politics. 

Two problems are identified throughout our analysis. First, there is no harmonized approach across the two social media platforms. Second, the scope and vision of both SMCs differ significantly when it comes to geographical variations in their take-down operations. While we observe parallel trends in terms of their interest in regime characteristics, defining state-linked networks, detecting those networks, and establishing the main premises of their operations to take down these networks require a more coherent approach. As a part of such policy-relevant outlooks, transparency concerns are also important. Meta, for example, does not provide user-level meta data, making further analysis difficult. More data will enable the use of more fruitful analyses, such as comparisons within and between countries. 

However, we also note that a possible problem for external validity arises with the existence of other US or China-based social media platforms. YouTube, TikTok and many other social media environments have a content moderation policy. Assuming they made their state-linked operations public, we would know more about the political nature of SMCs' take-down operations. This would be more interesting with TikTok, which has been under pressure by US authorities as a Chinese social media corporation. This could cause a potential bias in terms of how we evaluate the effectiveness of SMCs' take-down operations. We need to consider the possibility that these platforms have different policies, which could impact our findings. This could lead to different conclusions about the political nature of SMCs' take-down operations. Therefore, further studies can change our findings or generate more robustness. However, a level of transparency is needed to deal with the problem of SMCs' reaction to state-linked information operations.

\bibliography{main}

\newpage
\setcounter{figure}{0}
\renewcommand{\thefigure}{SI-\arabic{figure}}
\setcounter{table}{0}
\renewcommand{\thetable}{SI-\arabic{table}}

\FloatBarrier
\appendix
\section{Supplamentary Information}

\subsection*{Compiling Data for Meta and Twitter's Take Down Operations}

We compiled our data from Twitter and Meta's announcements. In each news, they announced country of origin, targeted countries, modality of operation and many other information such as domestic and foreign elements. We curated a dataset for reports between 2018 and 2022. During this time, they targeted different countries for state-linked operations. Table \ref{table:stats} shows all countries that Meta and Twitter focused on regarding different information operations. For instance, Russia originated operations identified by Meta and Twitter for 21 and 9 times, respectively for state-linked operations. 

Meta carry out its take-down operations against accounts originating in 48 distinct countries. We only coded state-linked operations if Meta clearly emphasized any linkage with governments, military, political parties and municipalities. 
This provided more insight to compare Meta with Twitter's take-down operations, since Twitter -from the beginning- announced state-linked operations. In the case of Twitter, Russia can set up networks in other countries such as Ghana and Nigeria. In such cases, we did not coded Ghana and Nigeria as being targeted by Twitter's counter operations.\footnote{\url{https://www.theguardian.com/technology/2020/mar/13/facebook-uncovers-russian-led-troll-network-based-in-west-africa}} 
We should also note that Twitter make user-level data public, while Meta does not provide such data. 
This is a significant difference between these SMCs. User level data is significant to understand the inherent mechanism of disinformation networks.

Regarding regression sample, we excluded some countries because of lack of data. For instance, we cannot retrieve GDP Per Capita data for Eritrea, South Sudan and Venezuela. Therefore, software automatically omits row with missing variables. 

Table \ref{table:stats} shows which countries are focused by Meta and Twitter. We observe that Meta was more active than Twitter in taking action against state-linked operations. There are 48 unique countries for Meta and this number is 22 for Twitter. 

\begin{table}[H]
\centering
\caption{Country Summary of Meta and Twitter's Geographic Response (2018-2021)}
\label{table:stats}
\begin{adjustbox}{width=0.8\textwidth}
\begin{tabular}{ |p{4cm}||p{4cm}|p{4cm}| }
\hline
\multicolumn{3}{|c|}{Country List} \\
\hline
Country & Number of Meta-Target & Number of Twitter-Target \\
\hline
Russia & 21 & 9 \\
Iran & 12 & 7\\
Mexico & 8 & 1 \\
Myanmar & 8 & - \\
Brazil & 4 & - \\
China & 4 & 6 \\
Georgia & 4 & - \\
Kyrgyzstan & 3 & - \\
Pakistan & 3 & - \\
WB/Gaza & 2 & - \\
El Salvador & 2 & - \\
Peru & 2 & - \\
Spain & 2 & 2 \\
Azerbaijan & 2 & - \\
Morocco & 2 & - \\
Ukraine & 2 & -\\
Nigeria & 1 & 1 \\
Nicaragua & 1 & - \\
Romania & 1 & - \\
Philippines & 1 & - \\
Saudi Arabia & 1 & 4 \\
Serbia (Yugoslavia) & 1 & 1 \\
Sudan & 1 & - \\
Thailand & 1 & 1 \\
Uganda & 1 & 1 \\
United States & 1 & - \\
Moldova & 1 & - \\
Algeria & 1 & - \\
Malaysia & 1 & - \\
KRG & 1 & - \\
Azerbeijan & 1 & - \\
Bangladesh & 1 & 1 \\
Belarus & 1 & - \\
Benin & 1 & - \\
Bolivia & 1 & - \\
Comoros & 1 & - \\
Costa Rica & 1 & - \\
Cuba & 1 & 1 \\
DRC & 1 & - \\
Ecuador & 1 & 1 \\
Egypt & 1 & 2 \\
Ethiopia & 1 & - \\
France & 1 & - \\
Ghana & - & 1 \\
Honduras & 1 & 1 \\
India & 1 & - \\
Indonesia & 1 & 1 \\
Israel & 1 & - \\
Jordan & 1 & - \\
Kazakhstan & 1 & - \\
Armenia & - & 1 \\
Tanzania & - & 1 \\
Turkey & - & 1 \\
United Arab Emirates & - & 1 \\
Venezuela & - & 4 \\
\hline
\end{tabular}
\end{adjustbox}
\end{table}

\newpage
\subsection*{Robustness analysis for regression models}

\begin{table}[H]
\centering
\caption{Regression Results (Linear Probability Modes-Pooled Time Series)}
\begin{adjustbox}{width=1.1\textwidth}
\begin{tabular}{l*{10}{c}}
\hline\hline
                    &\multicolumn{1}{c}{(1)}&\multicolumn{1}{c}{(2)}&\multicolumn{1}{c}{(3)}&\multicolumn{1}{c}{(4)}&\multicolumn{1}{c}{(5)}&\multicolumn{1}{c}{(6)}&\multicolumn{1}{c}{(7)}&\multicolumn{1}{c}{(8)}&\multicolumn{1}{c}{(9)}&\multicolumn{1}{c}{(10)}\\
                    &\multicolumn{1}{c}{Meta}&\multicolumn{1}{c}{Twitter}&\multicolumn{1}{c}{Meta}&\multicolumn{1}{c}{Twitter}&\multicolumn{1}{c}{Meta}&\multicolumn{1}{c}{Twitter}&\multicolumn{1}{c}{Meta}&\multicolumn{1}{c}{Twitter}&\multicolumn{1}{c}{Meta}&\multicolumn{1}{c}{Twitter}\\
                    &        b/se   &        b/se   &        b/se   &        b/se   &        b/se   &        b/se   &        b/se   &        b/se   &        b/se   &        b/se   \\
\hline
V-Dem Polyarchy     &      -0.402***&      -0.322***&      -0.228   &      -0.359** &      -0.327*  &      -0.359** &      -0.025   &       0.145   &      -0.237   &      -0.327** \\
                    &      (0.13)   &      (0.10)   &      (0.17)   &      (0.14)   &      (0.17)   &      (0.14)   &      (0.39)   &      (0.27)   &      (0.17)   &      (0.14)   \\
PVE                 &               &               &      -0.080*  &       0.017   &      -0.035   &       0.035   &      -0.035   &       0.033   &       0.206***&       0.120** \\
                    &               &               &      (0.04)   &      (0.03)   &      (0.05)   &      (0.04)   &      (0.05)   &      (0.04)   &      (0.07)   &      (0.05)   \\
UNGA Voting (w/USA) &               &               &               &               &       0.050   &      -0.245*  &       0.760   &       0.942   &       0.037   &      -0.249   \\
                    &               &               &               &               &      (0.28)   &      (0.15)   &      (0.87)   &      (0.62)   &      (0.23)   &      (0.15)   \\
GDP Per Capita (log)&               &               &               &               &       0.038   &       0.049** &       0.047   &       0.063***&       0.057** &       0.056** \\
                    &               &               &               &               &      (0.03)   &      (0.02)   &      (0.03)   &      (0.02)   &      (0.03)   &      (0.02)   \\
Population (log)    &               &               &               &               &       0.092***&       0.063***&       0.095***&       0.068***&       0.096***&       0.064***\\
                    &               &               &               &               &      (0.02)   &      (0.02)   &      (0.02)   &      (0.02)   &      (0.02)   &      (0.02)   \\
V-Dem Polyarchy $\times$ UNGA Voting (w/USA)&               &               &               &               &               &               &      -1.064   &      -1.778** &               &               \\
                    &               &               &               &               &               &               &      (1.20)   &      (0.86)   &               &               \\
PVE $\times$ UNGA Voting (w/USA)&               &               &               &               &               &               &               &               &      -0.873***&      -0.312** \\
                    &               &               &               &               &               &               &               &               &      (0.18)   &      (0.14)   \\
Constant            &       0.474***&       0.283***&       0.367***&       0.306***&      -1.399***&      -1.048***&      -1.705***&      -1.559***&      -1.628***&      -1.130***\\
                    &      (0.08)   &      (0.07)   &      (0.10)   &      (0.09)   &      (0.36)   &      (0.28)   &      (0.53)   &      (0.39)   &      (0.36)   &      (0.28)   \\
\hline
Number of clusters  &     168   &     168   &     168  &     168   &     168   &     168   &     168   &     168   &     168   &     168   \\
Number of observations&     672   &     672   &     672   &     672   &     672   &     672   &     672   &     672   &     672   &     672   \\
R-squared           &       0.053   &       0.065   &       0.072   &       0.067   &       0.197   &       0.190   &       0.201   &       0.212   &       0.253   &       0.204   \\
\hline\hline
\end{tabular}
\end{adjustbox}
\end{table}

\begin{table}[H]
\centering
\caption{Regression Results (Linear Probability Modes-Corssection)}
\begin{adjustbox}{width=1.1\textwidth}
\begin{tabular}{l*{10}{c}}
\hline\hline
                    &\multicolumn{1}{c}{(1)}&\multicolumn{1}{c}{(2)}&\multicolumn{1}{c}{(3)}&\multicolumn{1}{c}{(4)}&\multicolumn{1}{c}{(5)}&\multicolumn{1}{c}{(6)}&\multicolumn{1}{c}{(7)}&\multicolumn{1}{c}{(8)}&\multicolumn{1}{c}{(9)}&\multicolumn{1}{c}{(10)}\\
                    &\multicolumn{1}{c}{Meta}&\multicolumn{1}{c}{Twitter}&\multicolumn{1}{c}{Meta}&\multicolumn{1}{c}{Twitter}&\multicolumn{1}{c}{Meta}&\multicolumn{1}{c}{Twitter}&\multicolumn{1}{c}{Meta}&\multicolumn{1}{c}{Twitter}&\multicolumn{1}{c}{Meta}&\multicolumn{1}{c}{Twitter}\\
                  
\hline
V-Dem Polyarchy     &      -0.372***&      -0.303***&      -0.179   &      -0.290*  &      -0.322*  &      -0.288*  &       0.080   &       0.347   &      -0.226   &      -0.249   \\
                    &      (0.13)   &      (0.10)   &      (0.17)   &      (0.15)   &      (0.18)   &      (0.16)   &      (0.44)   &      (0.32)   &      (0.18)   &      (0.16)   \\
PVE                 &               &               &      -0.086*  &      -0.006   &      -0.039   &       0.028   &      -0.040   &       0.027   &       0.215***&       0.133** \\
                    &               &               &      (0.04)   &      (0.03)   &      (0.06)   &      (0.04)   &      (0.06)   &      (0.04)   &      (0.07)   &      (0.06)   \\
UNGA Voting (w/USA) &               &               &               &               &      -0.023   &      -0.292   &       0.909   &       1.185*  &      -0.030   &      -0.294   \\
                    &               &               &               &               &      (0.34)   &      (0.18)   &      (1.01)   &      (0.71)   &      (0.26)   &      (0.19)   \\
GDP Per Capita (log)&               &               &               &               &       0.037   &       0.048*  &       0.048   &       0.065** &       0.056*  &       0.056** \\
                    &               &               &               &               &      (0.03)   &      (0.02)   &      (0.03)   &      (0.03)   &      (0.03)   &      (0.02)   \\
Population (log)    &               &               &               &               &       0.091***&       0.074***&       0.095***&       0.079***&       0.097***&       0.076***\\
                    &               &               &               &               &      (0.02)   &      (0.02)   &      (0.02)   &      (0.02)   &      (0.02)   &      (0.02)   \\
V-Dem Polyarchy $\times$ UNGA Voting (w/USA)&               &               &               &               &               &               &      -1.411   &      -2.235** &               &               \\
                    &               &               &               &               &               &               &      (1.38)   &      (0.99)   &               &               \\
PVE $\times$ UNGA Voting (w/USA)&               &               &               &               &               &               &               &               &      -0.923***&      -0.378** \\
                    &               &               &               &               &               &               &               &               &      (0.21)   &      (0.17)   \\
Constant            &       0.454***&       0.285***&       0.337***&       0.277***&      -1.365***&      -1.221***&      -1.761***&      -1.847***&      -1.614***&      -1.323***\\
                    &      (0.08)   &      (0.07)   &      (0.10)   &      (0.09)   &      (0.37)   &      (0.30)   &      (0.58)   &      (0.43)   &      (0.38)   &      (0.31)   \\
\hline
Number of clusters  &               &               &               &               &               &               &               &               &               &               \\
Number of observations&     173   &     173   &     172   &     172  &     166   &     166  &     166   &     166   &     166   &     166   \\
\hline\hline
\end{tabular}
\end{adjustbox}
\end{table}

\begin{table}[H]
\centering
\caption{Regression Results (Logit Estimates, Cross-section, with UNGA Voting with China)}
\begin{adjustbox}{width=1.1\textwidth}
\begin{tabular}{l*{10}{c}}
\hline\hline
                    &\multicolumn{1}{c}{(1)}&\multicolumn{1}{c}{(2)}&\multicolumn{1}{c}{(3)}&\multicolumn{1}{c}{(4)}&\multicolumn{1}{c}{(5)}&\multicolumn{1}{c}{(6)}&\multicolumn{1}{c}{(7)}&\multicolumn{1}{c}{(8)}&\multicolumn{1}{c}{(9)}&\multicolumn{1}{c}{(10)}\\
                    &\multicolumn{1}{c}{Meta}&\multicolumn{1}{c}{Twitter}&\multicolumn{1}{c}{Meta}&\multicolumn{1}{c}{Twitter}&\multicolumn{1}{c}{Meta}&\multicolumn{1}{c}{Twitter}&\multicolumn{1}{c}{Meta}&\multicolumn{1}{c}{Twitter}&\multicolumn{1}{c}{Meta}&\multicolumn{1}{c}{Twitter}\\
                    &        b/se   &        b/se   &        b/se   &        b/se   &        b/se   &        b/se   &        b/se   &        b/se   &        b/se   &        b/se   \\
\hline
main                &               &               &               &               &               &               &               &               &               &               \\
V-Dem Polyarchy     &      -2.271***&      -3.601***&      -1.368   &      -3.691***&      -2.518** &      -3.561** &      -7.294   &     -28.594** &      -1.764   &      -2.921*  \\
                    &      (0.73)   &      (1.21)   &      (0.92)   &      (1.35)   &      (1.15)   &      (1.76)   &      (8.62)   &     (12.95)   &      (1.14)   &      (1.76)   \\
PVE                 &               &               &      -0.460** &       0.059   &      -0.184   &       0.176   &      -0.181   &       0.291   &      -7.722** &      -5.271***\\
                    &               &               &      (0.22)   &      (0.22)   &      (0.29)   &      (0.41)   &      (0.30)   &      (0.43)   &      (3.11)   &      (2.02)   \\
UNGA Voting Similarity with China&               &               &               &               &      -1.540   &       4.027   &      -5.386   &     -15.116   &      -0.350   &       5.208   \\
                    &               &               &               &               &      (2.50)   &      (3.31)   &      (7.48)   &      (9.29)   &      (2.57)   &      (3.29)   \\
GDP Per Capita (log)&               &               &               &               &       0.185   &       0.613** &       0.217   &       0.694** &       0.285   &       0.649** \\
                    &               &               &               &               &      (0.18)   &      (0.31)   &      (0.20)   &      (0.30)   &      (0.19)   &      (0.31)   \\
Population (log)    &               &               &               &               &       0.671***&       0.824***&       0.689***&       0.910***&       0.746***&       0.859***\\
                    &               &               &               &               &      (0.16)   &      (0.28)   &      (0.16)   &      (0.27)   &      (0.17)   &      (0.29)   \\
V-Dem Polyarchy $\times$ UNGA Voting Similarity with China&               &               &               &               &               &               &       5.735   &      29.964** &               &               \\
                    &               &               &               &               &               &               &     (10.26)   &     (15.15)   &               &               \\
PVE $\times$ UNGA Voting Similarity with China&               &               &               &               &               &               &               &               &       9.043** &       6.516***\\
                    &               &               &               &               &               &               &               &               &      (3.58)   &      (2.40)   \\
Constant            &       0.122   &      -0.455   &      -0.473   &      -0.392   &     -11.320***&     -22.808***&      -8.626   &      -8.642   &     -14.481***&     -24.842***\\
                    &      (0.39)   &      (0.51)   &      (0.50)   &      (0.61)   &      (4.25)   &      (6.76)   &      (6.09)   &      (8.50)   &      (4.42)   &      (6.54)   \\
\hline
Number of clusters  &               &               &               &               &               &               &               &               &               &               \\
Number of observations&     168   &     168   &     168   &     168   &     168   &     168   &     168   &     168   &     168   &     168   \\
\hline\hline
\end{tabular}
\end{adjustbox}
\end{table}

\begin{figure}
\centering
\includegraphics[width=\linewidth]{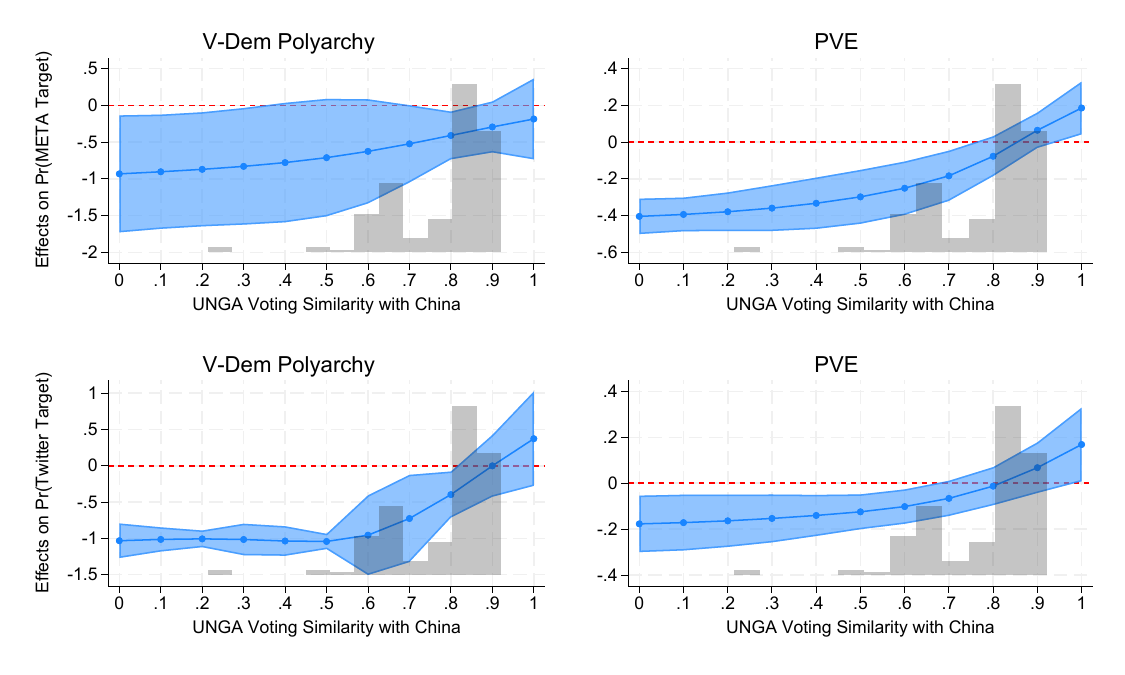}
\caption{Conditional Effects of Polyarchy and Political Stability on Meta and Twitter's Country Targeting}
\end{figure}

\begin{table}[H]
\centering
\caption{Regression Results (Logit Estimates, Cross-section, with User Statistics)}
\begin{adjustbox}{width=1.1\textwidth}
\begin{tabular}{l*{10}{c}}
\hline\hline
                    &\multicolumn{1}{c}{(1)}&\multicolumn{1}{c}{(2)}&\multicolumn{1}{c}{(3)}&\multicolumn{1}{c}{(4)}&\multicolumn{1}{c}{(5)}&\multicolumn{1}{c}{(6)}&\multicolumn{1}{c}{(7)}&\multicolumn{1}{c}{(8)}&\multicolumn{1}{c}{(9)}&\multicolumn{1}{c}{(10)}\\
                    &\multicolumn{1}{c}{Meta}&\multicolumn{1}{c}{Twitter}&\multicolumn{1}{c}{Meta}&\multicolumn{1}{c}{Twitter}&\multicolumn{1}{c}{Meta}&\multicolumn{1}{c}{Twitter}&\multicolumn{1}{c}{Meta}&\multicolumn{1}{c}{Twitter}&\multicolumn{1}{c}{Meta}&\multicolumn{1}{c}{Twitter}\\
                    &        b/se   &        b/se   &        b/se   &        b/se   &        b/se   &        b/se   &        b/se   &        b/se   &        b/se   &        b/se   \\
\hline
main                &               &               &               &               &               &               &               &               &               &               \\
V-Dem Polyarchy     &      -2.271***&      -3.601***&      -1.368   &      -3.691***&      -1.943*  &      -2.916*  &      -2.883   &       4.142   &      -1.402   &      -2.277   \\
                    &      (0.73)   &      (1.21)   &      (0.92)   &      (1.35)   &      (1.07)   &      (1.58)   &      (2.39)   &      (4.09)   &      (1.06)   &      (1.56)   \\
PVE                 &               &               &      -0.460** &       0.059   &      -0.824** &      -0.503   &      -0.821** &      -0.519*  &       0.768   &       0.472   \\
                    &               &               &      (0.22)   &      (0.22)   &      (0.36)   &      (0.31)   &      (0.36)   &      (0.30)   &      (0.72)   &      (0.55)   \\
UNGA Voting (w/USA) &               &               &               &               &       2.547   &      -0.230   &       0.445   &      14.071*  &       2.256   &      -1.802   \\
                    &               &               &               &               &      (1.63)   &      (2.13)   &      (4.88)   &      (7.67)   &      (1.90)   &      (2.46)   \\
GDP Per Capita (log)&               &               &               &               &      -0.101   &       0.457   &      -0.135   &       0.606*  &       0.043   &       0.495   \\
                    &               &               &               &               &      (0.27)   &      (0.30)   &      (0.29)   &      (0.33)   &      (0.28)   &      (0.31)   \\
Facebook Users (\%Population)&               &               &               &               &       2.953** &               &       3.077** &               &       2.305   &               \\
                    &               &               &               &               &      (1.46)   &               &      (1.52)   &               &      (1.45)   &               \\
Twitter Users (\% Population)&               &               &               &               &               &       0.171   &               &      -0.194   &               &       0.181   \\
                    &               &               &               &               &               &      (1.49)   &               &      (1.51)   &               &      (1.47)   \\
V-Dem Polyarchy $\times$ UNGA Voting (w/USA)&               &               &               &               &               &               &       3.218   &     -24.207*  &               &               \\
                    &               &               &               &               &               &               &      (7.15)   &     (14.48)   &               &               \\
PVE $\times$ UNGA Voting (w/USA)&               &               &               &               &               &               &               &               &      -5.927** &      -3.586*  \\
                    &               &               &               &               &               &               &               &               &      (2.64)   &      (1.93)   \\
Constant            &       0.122   &      -0.455   &      -0.473   &      -0.392   &      -1.440   &      -4.843*  &      -0.636   &      -9.943** &      -2.411   &      -4.894*  \\
                    &      (0.39)   &      (0.51)   &      (0.50)   &      (0.61)   &      (1.96)   &      (2.49)   &      (2.72)   &      (3.87)   &      (2.13)   &      (2.54)   \\
\hline

Number of observations&     168   &     168   &     168   &     168   &     162  &     160   &     162   &     160   &     162   &     160   \\
\hline\hline
\end{tabular}
\end{adjustbox}
\end{table}

\begin{figure}[H]
\centering
\includegraphics[width=\linewidth]{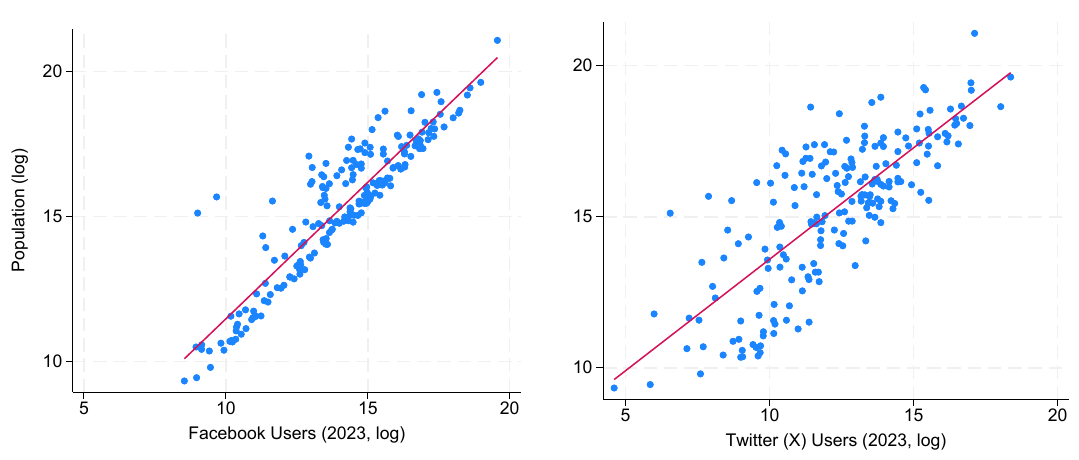}
\caption{User Statistics and Population. On the left, we plotted the correlation between user statistics and population. We captured a strong positive correlation between user statistics and population.}
\end{figure}

\begin{table}[H]

\caption{Correlation Table}
\begin{tabular}{lccc}
\hline
             & Population & Facebook Users'23 & Twitter Users'23 \\
\hline
Population   & 1.0000     &               &              \\
Facebook Users 2023 & 0.9126    & 1.0000        &              \\
Twitter Users 2023  & 0.4589    & 0.6350        & 1.0000       \\
\hline
\end{tabular}
\label{tab:correlation}
\end{table}

Table 6 statistically shows the association between user statistics and population. For Facebook, the correlation is .91, which is stronger than the association between Twitter's. Indeed, statistics with user statistics do not change our estimations as seen in Table 5. Since China, Iran and Russia do not have official user statistics for Facebook and Twitter, we replaced user statistics with population for a more comprehensive sampling. Our models with population and user statistics manifest similar results, as expected.

\begin{table}[H]
\centering
\caption{Variance inflation factors}
\begin{tabular}{l*{2}{c}}
\hline
            &Meta & Twitter\\
\hline
PVE         &      2.707492&    2.707492\\
V-Dem Polyarchy&      2.041271&    2.041271\\
GDP Per Capita (log)&              2.405718&    2.405718\\
UNGA Voting (w/USA)     &                 2.07146&     2.07146\\
Population(log)&               1.329778&    1.329778\\
\hline
\(N\)       &                166&         166\\
\hline
\end{tabular}
\end{table}

\begin{table}[H]
\centering
\begin{tabularx}{\textwidth}{|X|X|}
\hline
\textbf{Target of State-Linked Info-Ops (Meta)} & \textbf{Target of State-Linked Info-Ops (Twitter)} \\ \hline
United States(7), Qatar(7), Iran(6), United Kingdom(5), Yemen(5), Turkey(5), Morocco(4), Sudan(4), Lebanon(4), Egypt(3), Libya(3), Saudi Arabia(3), Syria(3), Tunisia(3), Ukraine(3), Jordan(3), Kazakhstan(2), Bahrain(2), Germany(2), India(2), Indonesia(2), Israel(2), Spain(2), Hong Kong(2), Comoros(2), Romania(1), Latvia(1), Estonia(1), Lithuania(1), Armenia(1), Azerbaijan(1), Georgia(1), Tajikistan(1), Uzbekistan(1), Moldova(1), Russia(1), Kyrgyzstan(1), Afghanistan(1), Albania(1), Algeria(1), France(1), Iraq(1), Mexico(1), Pakistan(1), Serbia(1), South Africa(1), Italy(1), Austria(1), Nigeria(1), Senegal(1), Togo(1), Angola(1), China(1), UAE(1), West Bank and Gaza(1), Somalia(1) & USA (5), Qatar (5), Hong Kong (4), Turkey (2), Uyghurs (2), Iran (1), Catalonia (1), Israel (1), Yemen (1), Spain (1), West Papua (1), Azerbaijan (1), NATO (1), EU (1), Central African Republic (1) \\ \hline
\end{tabularx}
\caption{Targets of State-Linked Information Networks Taken Down by Meta and Twitter}
\label{tab:Meta_twitter_info_networks}
\end{table}

We have identified that out of 196 operations where Meta took action to suspend or take down accounts, 116 were associated with state-linked activities. Within these 116 operations, 38 featured as involving inter-state state-linked information operations (the remaining networks deal with domestic propaganda). Similarly, for Twitter, all 47 datasets (as reported in separate announcements) are connected to state-linked activities. The targets of these state-linked operations encompass regional, domestic, and inter-state entities. Twitter explicitly lists the names of countries targeted in the second column of Table 8. This table provides a comprehensive overview of the countries, as well as regions, de facto states, and organizations, targeted by these state-linked information operations.

\end{document}